\documentclass[12pt]{article}

\usepackage{color}
\usepackage[dvips]{graphicx,psfrag}
\usepackage{amsmath,amssymb}
\usepackage{bm}

\baselineskip=\normalbaselineskip
\renewcommand{\baselinestretch}{1.34}
\renewcommand{\thefootnote}{\fnsymbol{footnote}}

\newcommand{\vev}[1]{\left\langle #1 \right\rangle}
\newcommand{\ket}[1]{\bigl|#1\bigr>}
\newcommand{\bra}[1]{\bigl<#1\bigr|}

\newcommand{\maru}[1]
{{\ooalign{\hfil#1\/\hfil\crcr\raise.167ex\hbox{\mathhexbox20D}}}}

\newcommand{\cO}{\mathcal{O}}

\newcommand{\bC}{\mathbb{C}}

\newcommand{\bZ}{\mathbb{Z}}

\newcommand{\del}{\partial}

\newcommand{\eq}[1]{(\ref{#1})}

\newcommand{\nn}{\nonumber}

\DeclareMathOperator{\str}{str}

\allowdisplaybreaks[1]

\newcommand{\gint}{\oint\hspace{-14pt}\bigcirc}

\newcommand{\ds}{\displaystyle}

\makeatletter
\@addtoreset{equation}{section}
\@addtoreset{theorem}{section}
\@addtoreset{definition}{section}
\@addtoreset{lemma}{section}
\@addtoreset{proposition}{section}

\begin{document}
\begin{flushright}
\parbox{40mm}{%
KUNS-2058 \\
{\tt hep-th/0701031} \\
January 2007}
\end{flushright}

\vfill

\begin{center}
{\Large{\bf 
Supermatrix models and multi ZZ-brane partition functions 
in minimal superstring theories
}}
\end{center}

\vfill

\begin{center}
{\large{Masafumi Fukuma}}\footnote%
{E-mail: {\tt fukuma@gauge.scphys.kyoto-u.ac.jp}} and  \!
{\large{Hirotaka Irie}}\footnote%
{E-mail: {\tt irie@gauge.scphys.kyoto-u.ac.jp}}   \\[2em]
Department of Physics, Kyoto University, 
Kyoto 606-8502, Japan \\

\end{center}
\vfill
\renewcommand{\thefootnote}{\arabic{footnote}}
\setcounter{footnote}{0}
\addtocounter{page}{1}

\begin{center}
{\bf abstract}
\end{center}

\begin{quote}
We study $(p,q)=(2,4k)$ minimal superstrings 
within the minimal superstring field theory 
constructed in hep-th/0611045. 
We explicitly give 
a solution to the $W_{1+\infty}$ constraints 
by using charged D-instanton operators, 
and show that the $(m,n)$-instanton sector 
with $m$ positive-charged and $n$ negative-charged ZZ-branes 
is described by an $(m+n)\times (m+n)$ supermatrix model. 
We argue that the supermatrix model 
can be regarded as an open string field theory 
on the multi ZZ-brane system. 
\end{quote}
\vfill
\renewcommand{\baselinestretch}{1.4}
\newpage

\section{Introduction}

Minimal noncritical (super)string theories 
\cite{Polyakov,KPZ,DHK,dsl}
are good toy models 
for investigating various aspects of string theory. 
They have fewer degrees of freedom 
but still share many features with their critical-string counterparts. 
Furthermore, there exists a string field theory 
\cite{fy1,fy2,fy3,FIS,fim,fi} 
which can completely describe 
both of fundamental strings (FZZT branes) 
and D-branes (D-instantons, ZZ branes), 
and has a clear relationship with Liouville-theory analysis 
\cite{DOZZ, fzz-t,zz,superLiouville,Mar,SeSh,KOPSS,Oku}. 
In particular, in \cite{fi} 
the spacetime which noncritical superstrings describe is clarified 
in terms of the two-component KP hierarchy. 

The aim of this letter is to further study 
the structure of spacetime in minimal superstring theories, 
especially the one emerging from 2-cut one-matrix models 
\cite{GroWit,PeShe,Napp,CDM,HMPN,ogu,BDJT,UniCom,SeSh2}. 
We show that spacetime probed by ZZ-branes 
has a description in terms of supermatrix models. 

This letter is organized as follows. 
In section 2, 
we make a brief review on type 0 minimal superstring theory 
and its string-field formulation \cite{fi}. 
In section 3, 
we explicitly give a solution to the $W_{1+\infty}$ constraints 
in the case of $\hat p=1$ minimal superstring theory,
and show that the partition function of the $(m,n)$-instanton sector 
with $m$ positive-charged and $n$ negative-charged ZZ-branes 
has a simple integral representation. 
In section 4, 
we show that the partition function of the $(m,n)$-instanton sector 
is expressed as an $(m+n)\times (m+n)$ hermitian supermatrix model, 
which can be interpreted as 
an open string field theory 
on the multi ZZ-brane system. 
Section 5 is devoted to discussions.

\section{Minimal superstring field theory}

Minimal type 0 superstring theory 
describes a product of 
minimal superconformal field theory (SCFT) 
and super Liouville field theory. 
Minimal SCFTs are characterized by the central charges
$
 \hat c^{\rm (matter)}(p,q)=1-2(q-p)^2/qp 
$
and are classified into two classes \cite{DiFrancesco:1988xz}:
\begin{itemize}
\item
\underline{{\bf even} minimal SCFT}: 
$(p,q)=(2\hat p, 2\hat q)$ with $\hat p+\hat q\in 2\bZ+1$
\item
\underline{{\bf odd} minimal SCFT}: \,
$(p,q)=(\hat p, \hat q)$ \,\,\,\, with 
$\hat p, \,\hat q$: odd
\end{itemize}
The scaling operators $\sigma^{[\mu]\,{\rm (matter)}}_n(z,\bar z)$ 
belonging to $(p,q)$ SCFT 
have conformal dimensions
\begin{align}
 \Delta_n^{[\mu] \,{\rm (matter)}}
  =\bar\Delta_n^{[\mu] \,{\rm (matter)}}
  =\frac{n^2-(\hat q-\hat p)^2}{8\hat q \hat p}
   +\frac{\mu}{16} 
 \quad 
 \Bigl(n\geq1; ~
 \mu=0\,(\mbox{NS-NS})~\mbox{or}~1\,(\mbox{R-R})
 \Bigr)\,,
\label{matter}
\end{align}
where $n$ and $\mu$ correspond to $(r,s)$ in the Kac table 
as $n=\hat q r-\hat p s$ and $\mu=r-s~(\mbox{mod}\,2)$. 
They are dressed with super Liouville field \cite{DHK} to become
\begin{align}
 \cO^{[\mu]}_n = \int\!d^2z \,\sigma^{[\mu]\,{\rm (matter)}}_n(z,\bar z)\,
  \sigma^{[\mu]\,{\rm (Liouville)}}_n(z,\bar z)
 \qquad(n\geq 1;~ \mu=0,1).  \label{OpLiou}
\end{align}

The partition function with R-R background flux $\nu$ 
\begin{align}
 \tau_\nu(x)\equiv \vev{ 
  \exp\Bigl[ (1/g)\sum_{n\geq1}\bigl(x^{[0]}_n \cO^{[0]}_n
    +x^{[1]}_n \cO^{[1]}_n \bigr)\Bigr] }_{\!\!\! \nu}
\end{align}
is given by a $\tau$ function of two-component KP (2cKP) hierarchy 
\cite{fi}. 
To explain this, we make a few preparations 
(see \cite{fi} for further explanations).

First we introduce 
two sets of chiral fermions on complex $\lambda$ plane,
\begin{align}
 &\psi^{(i)}(\lambda)
   =\sum_{r\in\bZ+1/2}\psi^{(i)}_r\lambda^{-r-1/2}
  ,\quad
  \bar\psi^{(i)}(\lambda)
   =\sum_{r\in\bZ+1/2}\bar\psi^{(i)}_r\lambda^{-r-1/2}  
  \quad (i=1,2)\,,\\
 & \bigl\{ \psi^{(i)}_r, \psi^{(j)}_s\bigr\}
   =\delta^{ij}\,\delta_{r+s,0}\,,
\end{align}
with the Dirac vacuum $\ket 0$, 
$\psi^{(i)}_r\,\ket 0=\bar\psi^{(i)}_r\,\ket 0=0$ $(r>0)$. 
We bosonize them as
$
 \psi^{(i)}(\lambda) = e^{\phi^{(i)}(\lambda)}
$,
$
 \bar\psi^{(i)}(\lambda)= e^{-\phi^{(i)}(\lambda)} 
$ 
$(i=1,2)$ 
with 
\begin{align}
 \phi^{(i)}(\lambda)=q^{(i)} + \alpha^{(i)}_0\,\ln\lambda
  -\sum_{n\neq 0}\,\frac{\alpha^{(i)}_n}{n}\,\lambda^{-n}, \qquad
 \bigl[ \alpha^{(i)}_m, \alpha^{(j)}_n\bigr]
   =m\,\delta^{ij}\,\delta_{m+n,0}.
\end{align}
The state 
$\ket{\nu}\equiv e^{\nu\,\bigl(q^{(1)}-q^{(2)}\bigr)}\,\ket{0}$ 
then describes the asymptotic state 
where the Fermi levels of the first $(i=1)$ and 
the second $(i=2)$ fermions differ by $2\nu$. 
This degree of freedom, $\nu$, can actually be interpreted 
as background R-R flux in the weak coupling region 
\cite{UniCom,SeSh2}.

We then introduce twisted bosons and twisted fermions 
on $\zeta\equiv\lambda^{\hat p} $ plane as
\begin{align}
 &\varphi^{(i)}_0(\zeta)\equiv 
  \phi^{(i)}(\lambda)
  ~~\Rightarrow~~
  \varphi^{(i)}_a(\zeta)\equiv \varphi^{(i)}_0(e^{2\pi ia}\zeta),\nn\\
 &c^{(i)}_a(\zeta)\equiv e^{\varphi^{(i)}_a(\zeta)},\quad
  \bar c^{(i)}_a(\zeta)\equiv e^{-\varphi^{(i)}_a(\zeta)}
  \quad (i=1,2;~a=0,1,\cdots,\hat p-1)\,,
\end{align}
from which the $W_{1+\infty}$ currents \cite{Winf} are defined as
\begin{align}
W^s (\zeta)
 &\equiv
    \sum_{n\in \mathbb Z} W_n^s \zeta^{-n-s}\nn\\
 &=s\,\sum_{a=0}^{\hat p-1}\, \Bigl(
     :\!\del^{s-1}c^{(1)}_a(\zeta)\cdot
     \bar c_a^{(1)}(\zeta)\!: 
    + (-1)^s 
     \Bigl[:\!\del^{s-1}c^{(2)}_a(\zeta)\cdot
     \bar c_a^{(2)}(\zeta)\!:\Bigr]_{\zeta \to -\zeta}\Bigr) \nn\\
 &=\sum_{a=0}^{\hat p-1}\,
   \bigl( 
     :e^{-\varphi^{(1)}_a(\zeta)}\del^s\,e^{\varphi^{(1)}_a(\zeta)}:
    +:e^{-\varphi^{(2)}_a(-\zeta)}\del^s\,e^{\varphi^{(2)}_a(-\zeta)}:
   \bigr).
\end{align}
The normal ordering $:\,\,:$ is taken 
with the ${\rm SL}(2,\bC)$ invariant vacuum on $\zeta$ plane, 
so that the monodromy like 
$\varphi^{(i)}_a(e^{2\pi i}\zeta)=\varphi^{(i)}_{a+1}(\zeta)$ 
should be interpreted as a relation to hold 
in correlation functions  
where $\bZ_{\hat p}$-twist fields are inserted 
at $\zeta=0$ and $\zeta=\infty$.

By introducing
\begin{align}
 \alpha^{[\mu]}_n\equiv \alpha^{(1)}_n + (-1)^\mu\,\alpha^{(2)}_n\,,
  \quad
  x^{[\mu]}_n\equiv \frac{1}{2}\,\bigl(
   x^{(1)}_n + (-1)^\mu x^{[\mu]}_n\bigr)
   \quad(\mu=0,1),
\end{align}
the partition function is expressed as
\begin{align}
 \tau_\nu(x)
  &= \bra{\nu\,}\, 
  e^{(1/g)\sum_{n\geq1}\bigl(x^{[0]}_n \alpha^{[0]}_n
    +x^{[1]}_n \alpha^{[1]}_n \bigr)} 
  \ket{\Phi}
  = \bra{\nu\,}\, 
  e^{(1/g)\sum_{n\geq1}\bigl(x^{(1)}_n \alpha^{(1)}_n
    +x^{(2)}_n \alpha^{(2)}_n \bigr)} 
  \ket{\Phi}\nn\\
 &\equiv \bra{x/g;\nu\,} \Phi\bigr>,
\end{align}
where the state $\ket\Phi$ satisfies the following two conditions \cite{fi}: 
\begin{align}
   \begin{array}{l}
    \bullet\,\,\mbox{\underline{decomposability}; 
     ~$\ket{\Phi}$ is written in the form 
      $e^{\rm (fermion~bilinear)}\,\ket{0}$} \\
    \bullet\,\,\mbox{\underline{$W_{1+\infty}$ constraints};\, 
      $W^s_n\,\ket{\Phi} = 0$ 
      ($s\geq 1;\, n\geq -s+1$)}
   \end{array} 
\label{conditions}
\end{align}
as in the bosonic case 
\cite{fkn1,dvv,fkn2,g,gn,KS,Krichever:1992sw,fkn3}.
The first condition is equivalent to the statement 
that $\tau_\nu(x)$ is a $\tau$ function 
of 2cKP hierarchy \cite{sato-sato,djkm,djkm2c,SW,KaLe}. 
The second one represents the whole set of 
the Schwinger-Dyson equations \cite{fkn2,g,KS,Krichever:1992sw,fkn3}. 
%
In the language of two-cut matrix models 
with symmetric double-well potentials, 
$\alpha^{[0]}_n$ 
({\em resp.} $\alpha^{[1]}_n$) describe 
symmetric ({\em resp.} antisymmetric) fluctuations of eigenvalues 
\cite{tt,newhat,UniCom},
so that $\alpha^{(1)}_n$ and $\alpha^{(2)}_n$ 
describe fluctuations in the right and the left well, respectively.

According to our ansatz on operator identification \cite{fi}, 
the excitations in the NS-NS and R-R sectors 
are collected into:
\begin{align}
 \mbox{\underline{NS-NS scalar}}:&\quad
  \del\varphi^{[0]}_0(\zeta)
  =\del\varphi^{(1)}_0(\zeta)+\del\varphi^{(2)}_0(\zeta)
  =\frac{1}{\hat p}\,\sum_{n\in\bZ}\,\alpha^{[0]}_n\,\zeta^{-n/\hat p-1},
\\
 \mbox{\underline{R-R scalar}}:&\quad
  \del\varphi^{[1]}_0(\zeta)
  =\del\varphi^{(1)}_0(\zeta)-\del\varphi^{(2)}_0(\zeta)
  =\frac{1}{\hat p}\,\sum_{n\in\bZ}\,\alpha^{[1]}_n\,\zeta^{-n/\hat p-1}.
\end{align}
Their connected correlation functions (or cumulants) 
in the presence of background R-R flux $\nu$ are given by
\begin{align}
 \vev{\del\varphi_0^{(i_1)}(\zeta_1)\cdots
  \del\varphi_0^{(i_N)}(\zeta_N)}_{\!\nu,{\rm c}}
 &=\biggl[\frac{\bra{x/g;\nu\,} 
   :\!\del\varphi^{(i_1)}_0(\zeta_1)\cdots
   \del\varphi^{(i_N)}_0(\zeta_N)\!:\ket{\Phi}}
   {\bra{x/g;\nu\,}\,\Phi\bigr>}\biggr]_{\rm c}\nn\\
 &=\sum_{h\geq0}\,g^{2h+N-2}\,
  \vev{\,\del\varphi^{(i_1)}_0(\zeta_1)\cdots
  \del\varphi^{(i_N)}_0(\zeta_N)\,}^{\!(h)}_{\!\nu,{\rm c}}.
\end{align}
Comparing the disk amplitudes with the algebraic curves 
of FZZT branes in super Liouville theory, 
we find the correspondence \cite{fi}:
\begin{align}
 \mbox{\underline{boundary states}}:\qquad
 \ket{{\rm\, FZZT}+;+\zeta}\Leftrightarrow 
  \varphi_0^{(1)}(\zeta), \quad
 \ket{{\rm\, FZZT}-;-\zeta}\Leftrightarrow
  \varphi_0^{(2)}(-\zeta)
\,.
\end{align}
Once a charged FZZT brane is located at a point 
in spacetime with coordinate $\zeta_{\rm bos}=\zeta^2$ \cite{fi}, 
it becomes a source of fundamental strings, 
with a bunch of worldsheets which are not connected with each other 
in the sense of worldsheet topology, 
but are connected in spacetime 
with their boundaries pinched at the same superspace point $\zeta$. 
These configurations are easily summed up to give 
an exponential form 
as in the bosonic case \cite{fy3}, 
realizing the spacetime combinatorics of Polchinski \cite{combi}:
\begin{align}
 \mbox{\underline{charged FZZT branes}}:\quad
  c^{(1)}_a(\zeta)= e^{\varphi^{(1)}_a(\zeta)},\quad
  c^{(2)}_a(-\zeta)= e^{\varphi^{(2)}_a(-\zeta)} \quad
 (a=0,1,\cdots,\hat p-1)\,.
\end{align}

As in the bosonic case \cite{fy1}, 
the D-instanton operators \cite{fi}
\begin{align}
 D_{ab}^{(ij)}=\gint\frac{d\zeta}{2\pi i}\,
  c_a^{(i)}(\zeta^{(i)})\bar c_b^{(j)}(\zeta^{(j)})
  =\gint\frac{d\zeta}{2\pi i}\,
  :\!e^{\varphi^{(i)}_a(\zeta^{(i)})-\varphi^{(j)}_b(\zeta^{(j)})}\!:
\nn\\
 \bigl(\mbox{$i=j$ with $a\neq b$; $i\neq j$ with $\forall(a,b)$};~
 \zeta^{(1)}\equiv +\zeta,\,\zeta^{(2)}\equiv-\zeta\bigr) \label{dinst}
\end{align}
commute with the $W_{1+\infty}$ generators:
\begin{align}
 \bigl[W^s_n,\,D_{ab}^{(ij)}\bigr]=0, \label{WD}
\end{align}
where the contour of \eq{dinst} surrounds 
$\zeta = \infty$ $\hat p$ times 
to resolve the monodromy of $\zeta$ plane. Equation \eq{WD} 
implies that given a state $\ket\Phi$ satisfying 
the $W_{1+\infty}$ constraints, 
one can construct another such state 
by multiplying it with $D^{(ij)}_{ab}$'s. 
By further requiring that the resulting state be decomposable 
$\bigl($
i.e.\ can be written as 
$e^{{\rm (fermion~bilinear)}}\,\ket{0}\,\bigr)$, 
the product of D-instanton operators 
must be accumulated to have the following form 
with fugacity
$\theta^{(ij)}_{ab}$ \cite{fy2}:
\begin{align}
 \ket{\Phi;\theta}\equiv 
  \prod_{i,j}\prod_{a,b}\,\exp\Bigl[
  \theta^{(ij)}_{ab}\,D^{(ij)}_{ab}\Bigr]\,
  \ket\Phi.
\end{align}
Note that $D^{(ij)}_{ab}$ $(i\neq j)$ 
includes the operator 
$e^{q^{(i)}-q^{(j)}}$   
and thus changes the relative Fermi levels. 
We thus have the following two classes of D-instanton operators:%
\footnote{
We have neglected the cocycles \cite{fi}
since their contributions can always be absorbed 
into $\theta^{(ij)}_{ab}$.
}

\noindent
\underline{neutral D-instanton operators $(a\neq b)$}:
\begin{align}
 D_{ab}^{(11)}&=\gint\frac{d\zeta}{2\pi i}\,
  c_a^{(1)}(\zeta)\,\bar c_b^{(1)}(\zeta)
  =\gint\frac{d\zeta}{2\pi i}\,
  :\!e^{\varphi^{(1)}_a(\zeta)-\varphi^{(1)}_b(\zeta)}\!: \,,\nn\\
 D_{ab}^{(22)}&=\gint\frac{d\zeta}{2\pi i}\,
  c_a^{(2)}(-\zeta)\,\bar c_b^{(2)}(-\zeta)
  =\gint\frac{d\zeta}{2\pi i}\,
  :\!e^{\varphi^{(2)}_a(-\zeta)-\varphi^{(2)}_b(-\zeta)}\!:.
\end{align}

\noindent
\underline{charged D-instanton operators $(\forall a,\forall b)$}:
\begin{align}
 D_{ab}^{(12)}&=\gint\frac{d\zeta}{2\pi i}\,
  c_a^{(1)}(\zeta)\,\bar c_b^{(2)}(-\zeta)
  =\gint\frac{d\zeta}{2\pi i}\,
  :\!e^{\varphi^{(1)}_a(\zeta)-\varphi^{(2)}_b(-\zeta)}\!: \,,\nn\\
 D_{ab}^{(21)}&=\gint\frac{d\zeta}{2\pi i}\,
  c_a^{(2)}(-\zeta)\,\bar c_b^{(1)}(\zeta)
  =\gint\frac{d\zeta}{2\pi i}\,
  :\!e^{\varphi^{(2)}_a(-\zeta)-\varphi^{(1)}_b(\zeta)}\!:.
\end{align}

In the weak coupling limit, $g\to+0$, 
one D-instanton amplitude $\bigl<D^{(ij)}_{ab}\bigr>$ can be evaluated 
by the saddle points $\zeta=\zeta_*$ of the exponent 
$\Gamma^{(ij)}_{ab}(\zeta)\equiv 
 \bigl<\varphi^{(i)}_a(\zeta^{(i)})\bigr>^{\!\!(h=0)}
   -\bigl<\varphi^{(j)}_b(\zeta^{(j)})\bigr>^{\!\!(h=0)}$ 
as $\bigl<D^{(ij)}_{ab}\bigr>\sim
  e^{(1/g)\,\Gamma^{(ij)}_{ab}(\zeta_*)}$ \cite{FIS,fim,fi}. 
This gives the relation between the amplitudes 
of the ZZ-branes and those of the FZZT-branes \cite{Mar} 
(see \cite{fi} for a detailed analysis).

\section{$\hat p =1$ minimal superstrings}

A great simplification occurs when $\hat p=1$, 
because $\zeta=\lambda$ in this case 
and the Dirac vacuum $\ket 0$ 
gives a trivial solution to the $W_{1+\infty}$ constraints. 
The general solutions are then given by
\begin{align}
 \ket{\Phi;\,\theta_+,\theta_-}
  =e^{\theta_+D_+}e^{\theta_-D_-}\,\ket{0},
\label{state}
\end{align}
where 
$ D_+\equiv D_{00}^{(12)} $, 
$ D_-\equiv D_{00}^{(21)} $,
and $\theta_+\equiv \theta_{00}^{(12)}$, 
$\theta_-\equiv \theta_{00}^{(21)}$. 
One can easily see 
that these states actually satisfy 
both of the conditions \eq{conditions}.
The fugacities $\theta_\pm$ represent the moduli of solutions. 
Note that there are no neutral ZZ-branes 
when $\hat p=1$.


The ground canonical partition function 
is expanded as 
\begin{align}
 \tau_\nu(x)= \bra{x/g; \nu\,}\, \Phi;\theta_+,\theta_-\bigr>
  =\sum_{m,\,n;\,\,m-n=\nu}\,
    \frac{\theta_+^m\, \theta_-^n}{m!\,n!}\,\,Z_{m,n}(x),
\end{align}
where
\begin{align}
Z_{m,n}(x)&= \bra{x/g;\nu=m-n}\,D_+^m\,D_-^n\,\ket{0} \nn\\
 &=\gint \prod_{r=1}^m  \frac{d\zeta^+_r}{2\pi i}\,
   \prod_{\alpha=1}^n \frac{d\zeta^-_\alpha}{2\pi i}
   \cdot (-1)^{(n+m)(n+m-1)/2}\,\times \nn\\
 &\qquad\times\bra{x/g; \nu=m-n\,}\,
  \prod_{r=1}^m
   :\!e^{\varphi^{(1)}_0(\zeta^+_r)-\varphi^{(2)}_0(-\zeta^+_r)}\!: 
  \prod_{\alpha=1}^n
   :\!e^{\varphi^{(2)}_0(-\zeta^-_\alpha)-\varphi^{(1)}_0(\zeta^-_\alpha)}\!: 
  \ket{0} \nn\\
\end{align}
is the partition function of the $(m,n)$-instanton sector 
with $m$ positive-charged and $n$ negative-charged ZZ-branes. 
This can be rewritten as 
\begin{align}
 Z_{m,n}(x)&=\bra{x/g; \nu=m-n\,}\,D_+^m\,D_-^n\,\ket{0}\nn\\
 &=\gint \prod_{r=1}^m  \frac{d\zeta^+_r}{2\pi i}\,
   \prod_{\alpha=1}^n \frac{d\zeta^-_\alpha}{2\pi i}\,\,
   \dfrac{ \ds
    \prod_{r<s}\bigl(\zeta^+_r-\zeta^+_s\bigr)^2\,
    \prod_{\alpha<\beta}\bigl(\zeta^-_\alpha-\zeta^-_\beta\bigr)^2
   }{ \ds 
    \prod_r\prod_\alpha \bigl(\zeta^+_r-\zeta^-_\alpha\bigr)^2
   }\,\cdot\,
   e^{(1/g)\bigl[
    \sum_r \Gamma(\zeta^+_r)\,-\,\sum_\alpha\Gamma(\zeta^-_\alpha)
    \bigr]} 
\label{Zmn},
\end{align}
where the D-instanton action $\Gamma(\zeta)$ is given by 
\begin{align}
\Gamma(\zeta)= \sum_{n=1}^{1+\hat q} (x_n^{(1)}+(-1)^{n+1}x_n^{(2)})\zeta^n 
\equiv  \sum_{n=1}^{1+\hat q} x_n\zeta^n.%
\end{align}%
\begin{figure}[htbp]
\begin{center}
\includegraphics[scale=0.9]{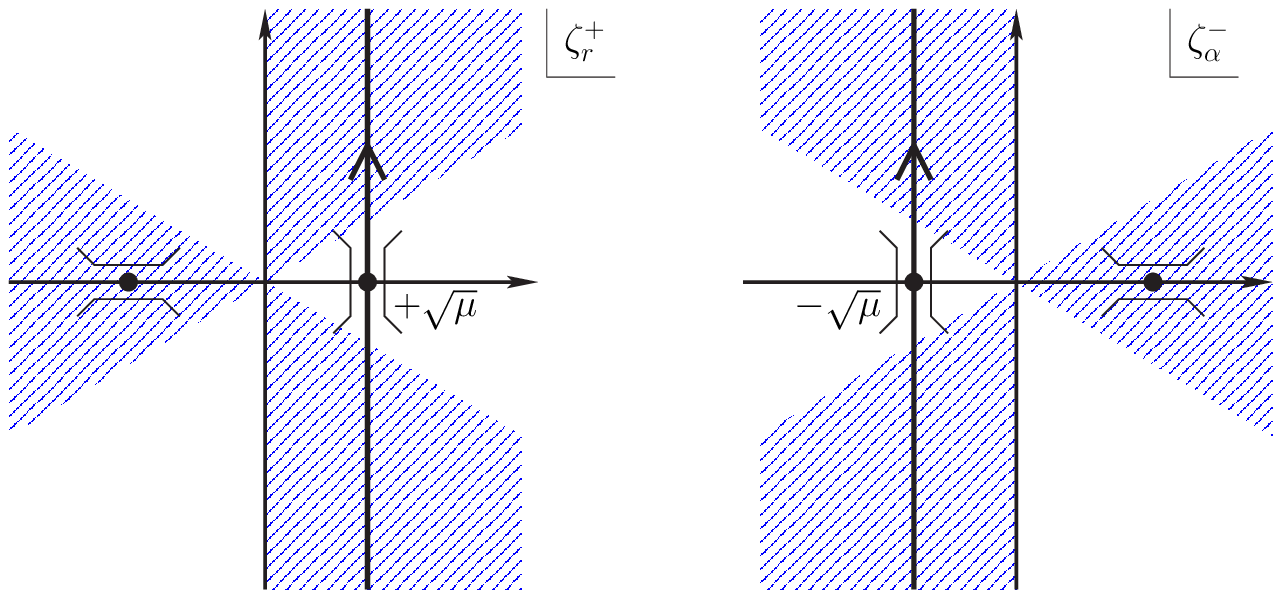}
\end{center}
\caption{\footnotesize{
Contours of $\zeta^+_r$ and $\zeta^-_\alpha$ 
for pure supergravity, $(p,q)=(2,4)$. 
The blobs are the saddle points of the functions 
$\Gamma(\zeta^+_r)$ and $\Gamma(\zeta^-_\alpha)$, 
and the accompanying bridges show their associated 
steepest descent directions. 
The shaded regions are the Stokes sectors 
where the real parts of the functions become negative for 
$|\zeta_r^+|,\, |\zeta_\alpha^-| \to \infty$.
}}
\label{contour}
\end{figure}%
The contours are chosen such that the state \eq{state} 
is defined well for some region in the parameter space 
of backgrounds, $\bigl\{x^{(i)}_n\bigr\}$.
In the case of pure supergravity, $(p,q)=(2,4)$ 
(or $(\hat p,\hat q)=(1,2)$), for example, 
one can take the contour as in fig.\ \ref{contour} 
if we take a background as
$x^{(1)}_3=x^{(2)}_3=1/3$ and $x^{(1)}_1=x^{(2)}_1=-\mu$. 
In fact, this background leads to 
$\Gamma(\zeta)=(2/3)\,\zeta^3-2\mu\,\zeta$,  
and as is investigated in \cite{fi}, 
when $\mu>0$ there exists a stable saddle point 
at $\zeta^+_{r\,*}=+\sqrt{\mu}$ for 
$\Gamma(\zeta^+_r)$ 
and at $\zeta^-_{\alpha\,*}=-\sqrt{\mu}$ 
for $-\Gamma(\zeta^-_\alpha)$, 
amounting to 
$\sum_r \Gamma(\zeta^+_{r\,*})
-\sum_\alpha\Gamma(\zeta^-_{\alpha\,*})
=-2(m+n)\,\mu^{3/2}/3<0$.  
This corresponds to the 1-cut phase of \cite{SeSh}. 
In fact, the string susceptibility 
$r^2\equiv -g^2\,(\del^2/\del\mu^2)\,\ln\tau_\nu$ 
with background R-R flux $\nu$ 
satisfies the string equation 
\cite{PeShe,Napp,CDM,HMPN,BDJT,UniCom,fi}
\begin{align}
 \mu\,r + \frac{1}{2}\,r^3
  - g^2\,\Bigl(\frac{1}{4}\,r'' + \frac{\nu^2}{r^3}\Bigr)=0, 
\end{align}
and has no perturbative parts when $\mu>0$. 
This implies that the partition function 
is fully described by stable D-instantons. 
The partition function in the other region ($\mu<0$) 
is then obtained by analytic continuation. 

\section{Supermatrix models}

The partition function of the $(m,n)$-instanton sector, 
\eq{Zmn}, can be further rewritten 
as an integration over $(m+n)\times (m+n)$ hermitian supermatrices:%
\footnote{
Here the term ``hermitian'' is in a formal sense  
as is the case in bosonic matrix integrals 
with unstable potentials. 
In fact, the eigenvalues are analytically continued 
into a complex plane as in fig.\ \ref{contour} 
in order to make the integral finite.
} 
\begin{align}
Z_{m,n}(x)=\int d\Phi \,e^{(1/g)\str\Gamma (\Phi)}\,.
\end{align}
Here $\Phi=\Phi^\dagger \in {\rm SMat}(m|n)$ and the measure $d\Phi$ is defined with 
\begin{align}
||d\Phi ||^2\equiv \str d\Phi^2\,.
\end{align}
In fact, assuming that this $\Phi$ is diagonalized as 
\begin{align}
 \Phi = V \Lambda V^{-1}=
   V\cdot
  \left(\begin{array}{cccccc}
   \zeta^+_1 &      &          &         &      &  \\
             &\ddots&          &         & 0    &  \\
             &      &\zeta^+_m &         &      &  \\
             &      &          &\zeta^-_1&      &  \\
             & 0    &          &         &\ddots&  \\
             &      &          &         &      &\zeta^-_n
  \end{array}\right)
  \cdot V^{-1}
\nn 
\end{align}
with $V\in U(m|n)$, one can rewrite the norm of the matrix as
\begin{align}
||d\Phi ||^2 &\equiv \str d\Phi^2=\str (d\Lambda^2+[d\Omega,\Lambda]^2) \nn\\
&= \sum_{r} (d\zeta_r^+)^2-\sum_{\alpha} (d\zeta_\alpha^-)^2 \,+ \nn\\
& \quad +2\,\sum_{r<s} (\zeta_r^+-\zeta_s^+)^2\,|d\Omega_{rs}|^2 
   -2\,\sum_{\alpha<\beta}(\zeta_\alpha^--\zeta_\beta^-)^2\,
    |d\Omega_{\alpha \beta}|^2\,
   +2\,\sum_{r,\alpha} (\zeta_r^+-\zeta_\alpha^-)^2\,
    |d\Omega_{r\alpha}|^2
\end{align}
with $d\Omega\equiv V^{-1}dV$. 
The measure can thus be factorized 
into those of eigenvalues $\{\zeta^+_r\}\cup \{\zeta^-_\alpha\}$ 
and angles $V\in U(n|m)$ as%
\footnote{
We have neglected contributions from $U(1)^{m+n}\subset U(m|n)$ as usual. 
Note that when $m=n$ 
the Jacobian can be collected 
into a single determinant 
due to the Cauchy identity:
\begin{align}
 d\Phi= dV\,\prod_{r=1}^n d\zeta^+_r\,
   \prod_{\alpha=1}^n d\zeta^-_\alpha\,\,
   \Bigl[{\det}_{r\alpha}
   \Bigl(\frac{1}{\zeta_r^+-\zeta_\alpha^-}\Bigr)\Bigr]^2. \nn
\end{align}
}
\begin{align}
 d\Phi= dV\,\prod_{r=1}^m d\zeta^+_r\,
   \prod_{\alpha=1}^n d\zeta^-_\alpha\,\,
   \dfrac{ \ds
    \prod_{r<s}\bigl(\zeta^+_r-\zeta^+_s\bigr)^2\,
    \prod_{\alpha<\beta}\bigl(\zeta^-_\alpha-\zeta^-_\beta\bigr)^2
   }{ \ds
    \prod_r\prod_\alpha \bigl(\zeta^+_r-\zeta^-_\alpha\bigr)^2
   }
\,,
\end{align} 
where $dV$ is the Haar measure for $U(m|n)$: 
$||dV||^2\equiv -{\rm str}\,\bigl(V^{-1}dV\bigr)^2$. 
The Jacobian correctly gives the factor in \eq{Zmn}.

\section{Discussions}

In this letter, we demonstrated that spacetime probed by ZZ-branes 
has a description in terms of supermatrix models.

This is another realization of the open/closed string duality. 
For the super Kazakov series, $(\hat p,\hat q)=(1,\hat q)$ 
(which includes $(p,q)=(2,4k)$ even minimal superstrings of section 2), 
a system of $m$ positive-charged and $n$ negative-charged ZZ-branes   
is described 
by a supermatrix $\Phi\in{\rm SMat}(m|n)$:
\begin{align}
 \Phi=\left( \begin{array}{cc}
   \Phi_{++} & \Phi_{+-} \\
   \Phi_{-+} & \Phi_{--} 
      \end{array}\right).
\end{align} 
The $m\times m$ matrix $\Phi_{++}$ describes open strings 
connecting $m$ positive-charged ZZ branes, 
while $n\times n$ matrix $\Phi_{--}$ describes open strings 
connecting $n$ negative-charged ZZ branes. 
Since these open strings connect two branes of the same charge, 
the resulting potential is repulsive, 
as can be seen from \eq{Zmn}.
On the other hand, 
the $m\times n$ (or $n\times m$) Grassmann-odd matrices 
$\Phi_{+-}$ (or $\Phi_{-+}$) 
describes open strings connecting oppositely charged ZZ-branes, 
so that the resulting potential turns out to be attractive.%
\footnote{
Although there are Grassmann-odd variables in supermatrix models, 
all the open strings connecting those ZZ-branes 
(with the same charge or with the opposite charges) 
are in the NS sector (see \cite{superLiouville}), 
since there exist only $\eta =-1$ (FZZT and ZZ) branes 
in our string field theory \cite{fi}. 
The same situations are noted in \cite{KOPSS,Oku}. 
}

An advantage of our string-field description 
of type 0 minimal superstrings 
is that such second-quantized picture is naturally obtained 
and summarized into a form of supermatrix model.%
\footnote{
See, e.g., \cite{SuperIntoMatrix} for former attempts 
to introduce supersymmetry into matrix models.
}
It should be interesting to investigate 
what roles these supervariables play 
in actual superstring theories 
and the corresponding matrix models. 

Note that this kind of supermatrix models  
do not need continuum limits. 
In this sense, they belong to a class of Kontsevich-type matrix models
\cite{witten,Kont}, 
and may have a possibility to describe 
the moduli space of super Riemann surfaces.


A further investigation of these matrix models 
and extension to more general $(p,q)$ cases 
are now in progress 
and will be reported in our future communication \cite{fi2}.

\section*{Acknowledgments}

We thank Tadashi Takayanagi for useful discussions. 
This work was supported in part by the Grant-in-Aid for 
the 21st Century COE
``Center for Diversity and Universality in Physics'' 
from the Ministry of Education, Culture, Sports, Science 
and Technology (MEXT) of Japan. 
MF and HI are also supported by the Grant-in-Aid for 
Scientific Research No.\ 15540269 and No.\ 18\textperiodcentered 2672, 
respectively, from MEXT.



\setlength{\itemsep}{5.\baselineskip}

\end{document}